\documentclass[journal]{vgtc}                
\ifpdf
  \pdfoutput=1\relax                   
  \usepackage{graphicx}                
  \DeclareGraphicsExtensions{.pdf,.png,.jpg,.jpeg} 
\else
  \usepackage{graphicx}                
  \DeclareGraphicsExtensions{.eps}     
\fi%

\graphicspath{{figures/}{pictures/}{images/}{./}} 

\usepackage{microtype}                 
\PassOptionsToPackage{warn}{textcomp}  
\usepackage{textcomp}                  
\usepackage{mathptmx}                  
\usepackage{times}                     
\usepackage{cite}                      
\usepackage{tabu}                      
\usepackage{booktabs}                  
\usepackage{mathrsfs}                  
\usepackage{amsmath}
\usepackage{wrapfig}
\usepackage{multirow}
\usepackage{amssymb}
\usepackage{pifont}
\newcommand{\cmark}{\ding{51}}%
\newcommand{\xmark}{\ding{55}}%
\usepackage[dvipsnames]{xcolor}
\usepackage[normalem]{ulem}
\usepackage{ulem}

\onlineid{0}

\vgtccategory{Research}
\vgtcpapertype{algorithm/technique}

\title{Visualizing Uncertainty in Probabilistic Graphs with \protect\\ Network Hypothetical Outcome Plots (NetHOPs)}

\author{
Dongping Zhang, 
Eytan Adar, and
Jessica Hullman
}
\authorfooter{
\item
 Dongping Zhang is with Northwestern University. E-mail: dzhang@u.northwestern.edu.
\item
 Eytan Adar is with the University of Michigan. E-mail:
 eadar@umich.edu.
 \item
 Jessica Hullman is with Northwestern University. E-mail:
 jhullman@northwestern.edu.
}

\shortauthortitle{Biv \MakeLowercase{\textit{et al.}}: Global Illumination for Fun and Profit}

\abstract{
Probabilistic graphs are challenging to visualize using the traditional node-link diagram. Encoding edge probability using visual variables like width or fuzziness makes it difficult for users of static network visualizations to estimate network statistics like densities, isolates, path lengths, or clustering under uncertainty. We introduce Network Hypothetical Outcome Plots (NetHOPs), a visualization technique that animates a sequence of network realizations sampled from a network distribution defined by probabilistic edges. NetHOPs employ an aggregation and anchoring algorithm used in dynamic and longitudinal graph drawing to parameterize layout stability for uncertainty estimation. We present a community matching algorithm to enable visualizing the uncertainty of cluster membership and community occurrence. We describe the results of a study in which 51 network experts used NetHOPs to complete a set of common visual analysis tasks and reported how they perceived network structures and properties subject to uncertainty. 
Participants' estimates fell, on average, within 11\% of the ground truth statistics, suggesting NetHOPs can be a reasonable approach for enabling network analysts to reason about multiple properties under uncertainty. Participants appeared to articulate the distribution of network statistics slightly more accurately when they could manipulate the layout anchoring and the animation speed. Based on these findings, we synthesize design recommendations for developing and using animated visualizations for probabilistic networks.
} 


\CCScatlist{ 
 \CCScat{K.6.1}{Management of Computing and Information Systems}%
{Project and People Management}{Life Cycle};
 \CCScat{K.7.m}{The Computing Profession}{Miscellaneous}{Ethics}
}

\teaser{
  \centering
  \includegraphics[width=\linewidth]{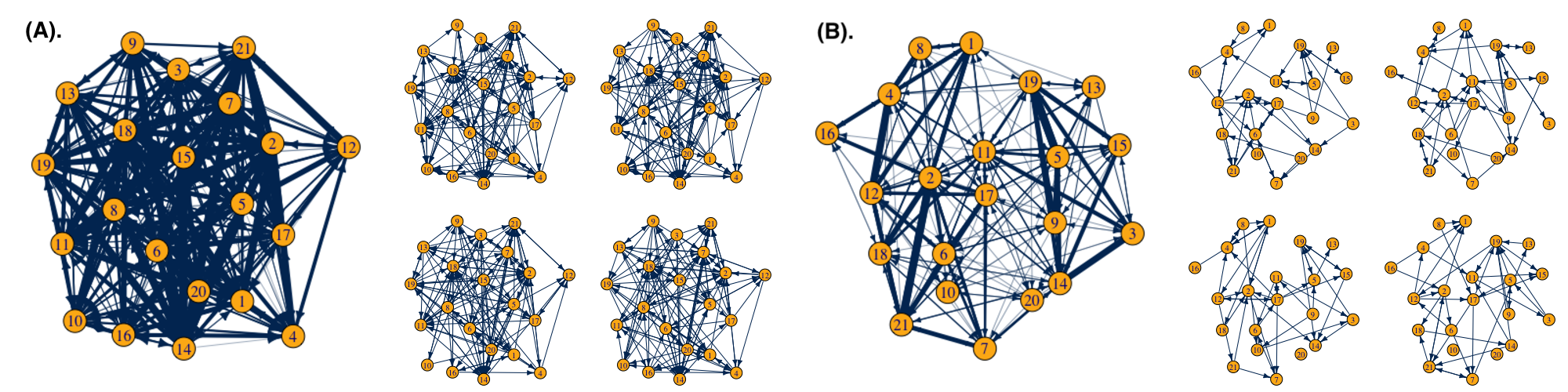}
  \caption{(A) and (B) are comparisons between static probabilistic networks and sampled network realizations from a random graph model defined by probabilistically weighted edges. Both (A) and (B) consist of 21 vertices with corresponding density values of 0.9 and 0.5.}
  \label{fig:teaser}
}

\vgtcinsertpkg

\begin{document}


\firstsection{Introduction}
\maketitle
Network data are prone to uncertainty. It is often unclear whether all relevant entities are included in a graph and if observed interactions are representative or occur merely by chance (e.g., \cite{kaiser2017ecosystem, young2019reconstruction}). For example, social network data are frequently collected through surveys, but it is well-known in the Social Network Analysis (SNA) community that network surveys are problematic due to selection bias, response bias, and missing responses (e.g., \cite{bell2007partner, wright2002sorry}). While technological affordances of online platforms reduce data uncertainty by making user interactions visible and persistent~\cite{gaver1991technology, leonardi2013enterprise, leonardi2018better}, uncertainty remains an issue when analysts binarize or predict social relations through statistical models based on the frequency of communication (e.g., \cite{adamic2003friends}). 

Uncertainty in the network analysis pipeline is sometimes addressed by imposing probabilities on edges as weights, resulting in a probabilistic graph. Although edge uncertainty can be visualized in a node-link diagram through visual encodings such as width, fuzziness, or grain~\cite{guo2015representing, maceachren2012visual}, the rendered graph is typically difficult to visually assess for practical use. For example, an analyst will likely find it challenging to respond to common graph analysis tasks, such as, ``What is the most likely shortest path length between node 16 and node 9?'', or ``What is the expected density?'' using either of the static visualizations shown in \autoref{fig:teaser}. In such cases, analysts must rely on a hard-to-decode visual channel not only to gain probability information about any single edge, which might be difficult to see due to the high density of a probabilistic graph (i.e., showing all edges with non-zero weights), but also to simultaneously integrate and process the joint probability from multiple edges for certain statistics (e.g., path lengths, isolates, and densities).

Similarly, although there are algorithms to identify clusters for static weighted networks (e.g., \cite{lu2014algorithms}), the uncertainty of community occurrence or cluster membership is difficult to visualize in a static diagram because traditional encodings (e.g., node coloring or convex hulls) are used to show deterministic community membership. These reasons may contribute to the relative lack of techniques available for visualizing probabilistic graphs to support exploratory network analysis.

To provide a solution, we introduce Network Hypothetical Outcome Plots (NetHOPs), a frequency-based uncertainty visualization technique that applies animated hypothetical outcomes~\cite{hullman2015hypothetical} to probabilistic graphs. NetHOPs dynamically visualize a set of realizations sampled from a network distribution defined by probabilistic edges. By showing independent possible realizations of a network, NetHOPs avoid the challenges of supporting judgments about network properties under uncertainty with static encodings. Instead, structures and properties can be estimated by attending to the temporal frequency of occurrence from a set of independent but equally representative realizations. 

A challenge arises when presenting network uncertainty as a sequence of node-link diagrams:  graph layout algorithms are intentionally optimized based on a static network. This induces a trade-off between more optimal visualization of each sampled realization and analysts' ability to preserve their mental maps~\cite{misue1995layout} of how vertices and edges relate across realizations. NetHOPs address this trade-off using an offline graph drawing approach and applying an aggregation and anchoring algorithm \cite{brandes2012visualization} to enable analysts to control layout stability and readability~\cite{brandes1997bayesian}. To address the challenges associated with visualizing community membership under uncertainty, NetHOPs employ community detection algorithms designed for static unweighted networks, for which we develop a community matching and coloring algorithm.

To demonstrate the use of NetHOPs in a realistic analysis setting and explore how well the technique supports uncertainty perception on networks, we contribute a user study in which 51 network experts used NetHOPs to complete a set of common visual analysis tasks, chosen based on network task taxonomies~\cite{lee2006task,ahn2013task}. 
The experts provided probability estimates and used distribution builders~\cite{sharpe2000distribution} to sketch their perceptions of uncertainty in network properties (e.g., density, shortest path, clusters). We find that NetHOPs allowed experts to visually assess the distributions of network structures fairly accurately: responses were within 11\% of the ground truth distributions (defined on the full set of visualized realizations) across tasks. When participants were instructed to adjust NetHOP parameters like the amount of layout stability, the animation speed, and other visual properties, we see some evidence of a small (7\%) additional average improvement in performance.
We present an exploratory analysis of our results that we use to reflect on how different visual network estimation tasks under uncertainty can be best supported. Based on our findings, we synthesize design recommendations for future development of NetHOPs and similar techniques as a solution to address network uncertainty.

\section{Background}
\subsection{Uncertainty Visualization}
Prior work demonstrates that people can make better decisions when uncertainty is effectively represented~\cite{joslyn2012uncertainty, nadav2009uncertainty, jung2015displayed, evans1997dynamic}, which makes uncertainty communication an important practice for scientific research \cite{taylor1994guidelines}. Empirical evidence suggests that framing probabilities as frequencies (e.g., 3/10 rather than 30\%) can make them easier to reason with~\cite{gigerenzer1995improve, hoffrage1998using}. Among other frequency-based uncertainty visualization techniques (e.g., \cite{garcia2010icon, kay2016ish, fernandes2018uncertainty}), hypothetical outcome plots (HOPs) \cite{hullman2015hypothetical,kale2018hypothetical} 
display a finite set of samples from a distribution as a series of animated frames. Some studies of simple 2D visualizations find that HOPs can lead to better estimates than error bars~\cite{hofman2020visualizing,hullman2015hypothetical,kale2018hypothetical} and other static techniques like static ensembles and violin plots~\cite{kale2018hypothetical}. 

We note several properties of HOPs that make them potentially interesting as a technique for displaying probabilistic graphs. First, HOPs are a natural choice for visualizing uncertainty when baseline visual encodings are already complex and hard to read, such that the addition of another encoding (e.g., width) or glyph (e.g., error bars), may not be effective. By using temporal frequency encoding, which perceptual research has found can be processed automatically without requiring counting~\cite{hasher1984automatic}, HOPs support intuitive estimation of event probabilities, which in a network application might include probabilities associated with structures of interest (e.g., edges, clusters, cliques).
Finally, by depicting multivariate samples, HOPs make it possible for a single visualization to show joint probabilities, which typically requires adding additional views. While our network model does not require joint probability depiction, other probabilistic networks can (e.g., ~\cite{hunter2008goodness}). 

\subsection{Graph Visualization and Animation}
Many network properties (e.g., centrality, clustering, density) have visual signatures, which make network visualization an important tool for network analysis. When well laid out, a node-link diagram can provide an effective bird's eye view of the network, making these diagrams a ubiquitous representation of graphs. Analysts use node-link diagrams to identify overall patterns such as cores and peripheries, communities or neighborhoods, and isolates and components, among others. While many structural configurations (e.g., reciprocity, triadic closure, or alternating path) can be summarized by network statistics or distributions \cite{lusher2013exponential}, analysts typically begin the analysis by examining structural dependencies and correlations from a comprehensive overview of the graph, which visualizations can help to provide.

Graph layout algorithms, like force-directed (spring embedder) layouts~\cite{eades1984heuristic}, stress minimization~\cite{brandes2008experimental,gansner2004graph}, and the stochastic gradient descent approach~\cite{zheng2018graph} can help to achieve aesthetic goals like minimizing edge crossing or overlapping nodes to support low-level visual tasks such as tracing each edge from source to target and seeing clusters~\cite{shneiderman2006network}.

Animated node-link diagrams are commonly used to visualize dynamic and longitudinal graphs~\cite{branke2001dynamic}, where each frame in the animation represents the graph at different states. It is possible but often impractical to show all graphs in a static visualization (e.g., small multiples~\cite{archambault2010animation}) due to large numbers of graphs. When designing network animations, it is insufficient to apply a layout algorithm to each state in a time-varying graph because the dynamic layout algorithm can produce inconsistent results when repeatedly applied to the same network data. Instead, layouts must be carefully engineered to help analysts preserve their mental maps of how the nodes and edges change over time \cite{misue1995layout, saffrey2008mental, purchase2008extremes}.

Layout computation for dynamic network visualizations can be categorized into offline computation based on the whole sequence of graphs \cite{brandes2011quantitative} and an online approach \cite{brandes2012visualization} that computes the graph layout one transition at a time \cite{frishman2008online,north1995incremental}. Layout stability can be addressed by anchoring vertices to fixed positions while animating edges \cite{moody2005dynamic} or by connecting vertices from different instances with external edges, which is conceptually similar to parallel-coordinates ~\cite{erten2003simultaneous, dwyer2006visual, dwyer2004visualising}. 

Animating random draws from a probabilistic graph model as applied in NetHOPs also requires addressing layout stability. However, instead of helping an analyst detect graph evolution over time, the goal of layout stability when animating a probabilistic graph is to facilitate more accurate probability estimates for network structures and make these entities of interest recognizable from frame to frame. NetHOPs' layout computation is based on the aggregation and anchoring approach, developed by Brandes et al. \cite{brandes2012visualization}, which is an instance of the offline drawing scenario where stress minimization is used to compute a reference layout based on all graphs in the sequence. 

\subsection{Visualizing Probabilistic Graphs}\label{sec:background-probabilistic-graph}
Probabilistic networks have a fixed vertex set with immutable vertex attributes. Edges are weighted, and weights are typically the probability of edge occurrence. Edge weights can be dependent by conditional probability or independent based on network context or network models used. From an analytical perspective, probabilistic networks have constrained usability because some common graph analysis tasks, such as cluster detection, are difficult to implement when edges are weighted by probability. Although there are existing algorithms to identify clusters and communities for a static weighted network (e.g., \cite{lu2013community, lu2014algorithms, jin2011center}), the uncertainty of community occurrence or cluster membership is impossible to visualize in a static network representation. 

Although visual variables \cite{bertin1983semiology} (e.g., width or fuzziness) can encode edge probability, this approach is unsuitable for denser probabilistic networks\cite{guo2015representing, maceachren2012visual}. This is because probabilistic graphs tend to have maximal connectivity, that is, all edges with non-zero weights must be present in the graph. Even if analysts can find relevant network configurations (e.g., path or alternating stars), they must decode and integrate the meaning of visual properties from multiple edges, which are often limited for supporting accurate visual judgments~\cite{hullman2015hypothetical,kosara2002useful, mackinlay1986automating}.

\begin{figure}[tb]
    \centering 
    \includegraphics[width=\columnwidth]{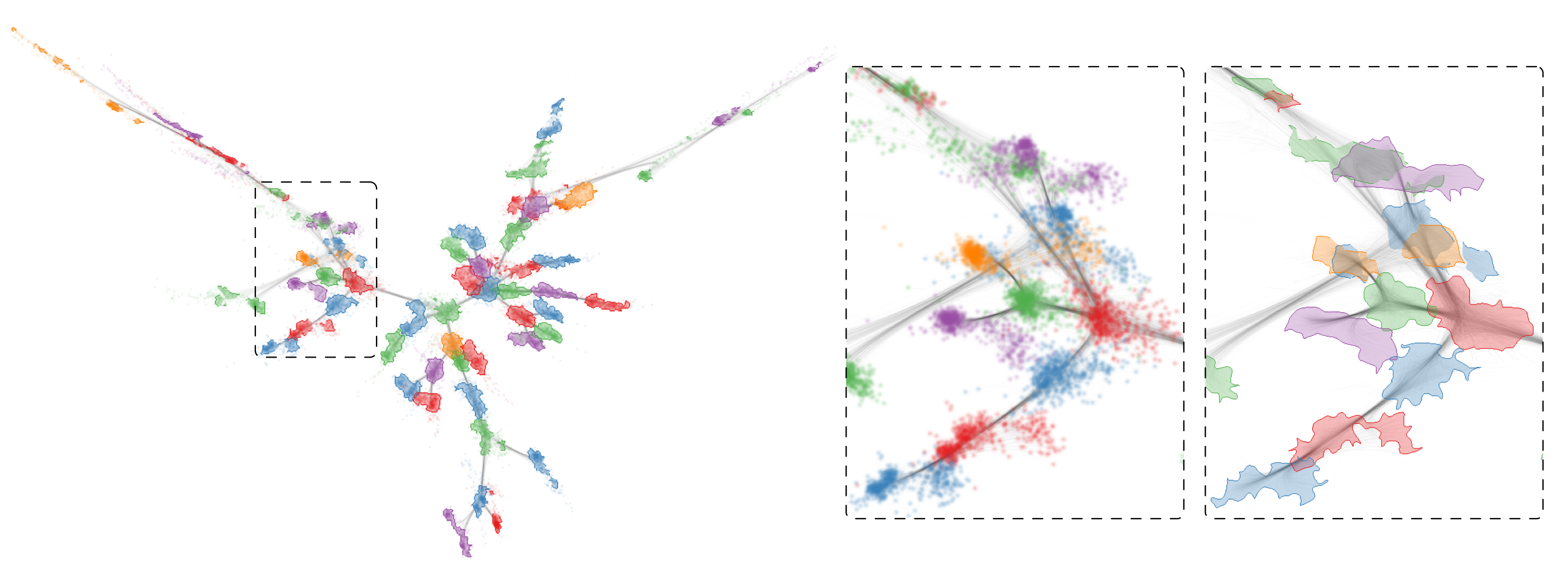}
    \setlength{\abovecaptionskip}{-5mm}
    \setlength{\belowcaptionskip}{-5mm}
    \caption{
    Probabilistic graph layout by Schulz et al. \cite{schulz2016probabilistic}.
    }
    \label{fig:prob-layout}
\end{figure}

One common strategy to simplify weighted graphs is to truncate or subset the edges by an analyst-defined threshold~\cite{romney1986culture, freeman1987words}. However, it is difficult to identify an optimal threshold that can maintain the structure of the networks without information loss. Perhaps because of these challenges, visualizing probabilistic networks using the node-link diagram is a relatively unexplored topic. One exception is Schulz et al. \cite{schulz2016probabilistic}, who developed a layout algorithm for probabilistic networks that can blend sampled realizations from a probabilistic graph model into a static visualization, shown in~\autoref{fig:prob-layout}. Their approach anchors and aligns different network realizations \cite{brandes2011quantitative} such that the same vertex can form a vertex cloud through node splatting \cite{holten2006hierarchical}. Edges are splatted and bundled \cite{holten2006hierarchical} to improve readability. However, improved readability of the graph as a whole comes at the cost of concealing edge uncertainty, such that it is difficult to imagine analysts estimating the distribution of properties like edge occurrence or path lengths. Conceptually, probabilistic graph layout and NetHOPs share many similarities, but the major difference is that probabilistic graph layouts aim to create a static visual representation through 2D graph embedding, such that many of the limitations for decoding distributional information from a static depiction of a probabilistic graph remain.

\subsection{Community Detection}
Graph clustering is an important topic in many network domains (e.g., \cite{bedi2016community, martinet2020robust, jia2012community, kim2014improved}). Network task taxonomies describe clustering and community identification as a class of topology-based tasks that is fundamental to network analysis and visualization \cite{ahn2013task, lee2006task}. 

Prior work proposes several robust community detection algorithms on static graphs based on modularity optimization (e.g., \cite{girvan2002community, newman2004finding}), Markov simulations \cite{van2000graph}, random walk \cite{pons2005computing}, and cluster structures \cite{rosvall2007information, rosvall2008maps}, to name a few. Many of these techniques are unsupervised or semi-supervised, so cluster boundaries and node memberships can be uncertain. Some interactive visualization tools communicate this uncertainty by showing crisp and overlapping community structures \cite{vehlow2013visualizing, wang2015ambiguityvis} to address fuzzy overlapping communities with different granularity.

When graphs are made dynamic through a temporal dimension, topology can change as vertices and edges appear and vanish. Hence, detected communities can emerge, merge, split, and disappear \cite{palla2007quantifying, rossetti2018community}. Most dynamic community detection approaches can be classified as instant optimal, temporal trade-off, or cross-time \cite{rossetti2018community}. The instant optimal approach searches communities individually using each graph in the sequence, and matches communities detected throughout the sequence (e.g., \cite{rosvall2010mapping, wang2008commtracker, hopcroft2004tracking}). We develop a variant of the instant optimal approach for NetHOPs so our technique can support cluster detection and community membership for probabilistic graphs.

\section{Network Hypothetical Outcome Plots (NetHOPs)}
NetHOPs animate a set of hypothetical outcomes sampled from a probabilistic graph model, designed to address the challenges associated with visualizing probabilistic networks. This representation removes the requirement that users visually integrate distributional information across static encodings of probability. With NetHOPs, analysts can perceive deterministic network statistics from each sampled realization (see \autoref{fig:teaser}), and identify edge probabilities, and probabilities of higher-level network structures, by integrating across the animated samples.

At a high level, the NetHOPs creation pipeline starts with formulating a probabilistic random graph model based on a given network dataset. The model provides a network data generating process enabling us to sample a sequence of different network realizations via a Monte Carlo process. We apply our instant-optimal community detection and matching algorithms (see \autoref{sec:design-community-matching}) to the network sequence so each individual realization is supplemented with additional measures that capture community structure across the set. We pass the network sequence to the visualization functions, which compute the layouts and use the additional community structure measures to color communities.

\subsection{Probabilistic Network Data}
NetHOPs are agnostic to how the probabilistic graph model is inferred as long as the graph model describes a network distribution and network sampling is enabled through a Monte Carlo process. In \autoref{sec:prototype}, we demonstrate the construction of a random graph model by treating edge occurrences as following a Bernoulli distribution parameterized by their corresponding probabilistic weights without dependency assumption. 

Our example is simple and aimed to convey the idea of probabilistic graphs. More rigorous network modeling approaches are available but are not the focus of our work. For example, efforts have been made to statistically quantify and model uncertainty in network edges by reconstructing the network through Bayesian inference \cite{young2019reconstruction, young2020bayesian}. The exponential-family random graph models (i.e., $p^*$ models) can also place conditional probabilities on user-specified structural configurations \cite{robins2007introduction, robins2007recent, lusher2013exponential, caimo2011bayesian}. Additionally, probabilistic networks can be easily created based on network context, such as the well-known Cognitive Social Structures (CSS) \cite{krackhardt1987cognitive} we use for demonstration.

Given a probabilistic graph model (e.g., a posterior distribution in a Bayesian network reconstruction pipeline~\cite{young2019reconstruction,young2020bayesian}), we can sample realizations through a Monte Carlo process, which generates a set $G$ of N graphs where $G^N = \{G^{1} = (V, E^{1}), \; \cdots, \; G^{n} = (V, E^{n})\}$. We can then supplement the sequence of realizations with additional information, including a stability score and community label for each vertex.

\subsection{Community Matching} \label{sec:design-community-matching}
NetHOPs are designed to support probabilistic versions of common visual network analysis tasks, such as estimating distributions of network statistics that are normally deterministic (e.g., shortest path, density, edge occurrence, isolates). Many of these can be supported primarily through layout engineering to preserve stability (see \autoref{sec:method-task}).

However, other properties, such as assessing the number of communities or the stability of community membership, require further optimizations. 
Cluster structures and membership can vary for each network realization that comprises NetHOPs due to small differences in configuration across realizations~\cite{rossetti2018community}. This problem is similar to the ``Ship of Theseus'' thought experiment~\cite{scaltsas1980ship} in which it is challenging to identify communities when their corresponding vertex memberships change partially or completely in different network realizations. 

\begin{figure}[tb]
    \centering 
    \includegraphics[width=\columnwidth]{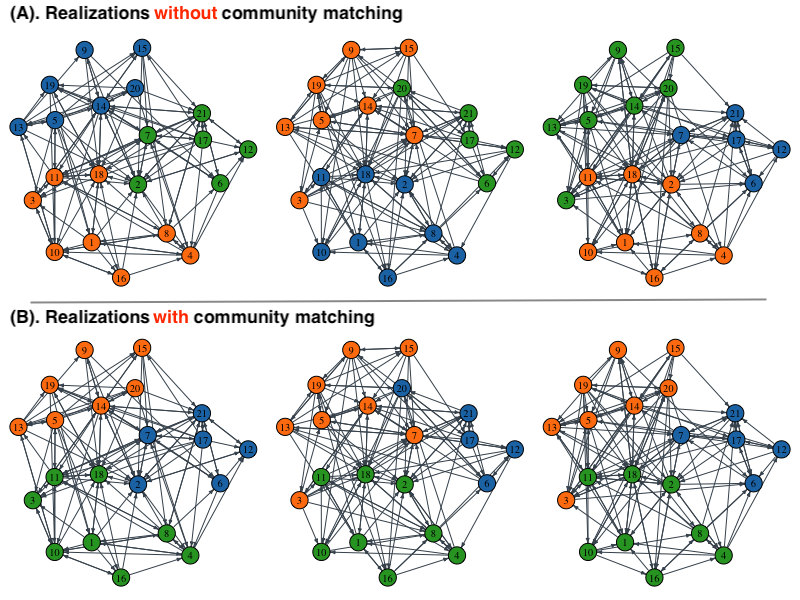}
    \setlength{\abovecaptionskip}{-5mm}
    \setlength{\belowcaptionskip}{-5mm}
    \caption{
    (A) Probabilistic network realizations with community structures detected by Brandes et al. \cite{brandes2007modularity}. Vertices are colored to indicate community membership. (B) Same as (A) but with a community matching algorithm applied as described in \autoref{sec:design-community-matching}}
    \label{fig:community-coloring}
\end{figure}

To demonstrate, the three static network visualizations in the first row of \autoref{fig:community-coloring} are network realizations sampled from a probabilistic graph derived from Krackardt's advice-seeking CSS dataset (see \autoref{sec:prototype} where we describe the data) positioned using the reference layout computed by aggregation. We apply a modularity-based community detection algorithm \cite{brandes2007modularity} to each of the network realizations, and colored vertices based on community membership without any analysis of relationships across realizations. All three realizations have three distinct communities, with similar vertex memberships. However, because community detection algorithms are typically not capable of leveraging cluster information across a set of related network realizations, communities will appear to change despite the fact that the clustering results are consistent. For example, the blue community in realization 2 has the same vertex membership as the orange community in realization 3, but we cannot identify these two communities as the same. 

To address this problem, we devise a two-step community matching algorithm. Given a sequence of N graphs $G^N$ where each $G^n = (V, E^n)$ with vertex set V and edge set $E^n = \{(i, j): \; i \neq j \; \text{and} \;i, j \in V\}$, a chosen community detection algorithm can assign a community label to each vertex as a vertex attribute $Y$ where $Y^N = \{Y^1_i,\cdots,Y^n_i | i \in V\}$.

We then use the community-labeled network realizations as inputs to a matching process based on the degree of vertices' community co-occurrences. This process starts with constructing a weighted full graph (which we call the ``co-community graph'') $G_F = (V, E_F)$. The weight matrix $W$ where $w_{i,j} = |\{Y^n_i = Y^n_j | n \in N\}|$ for $G_F$ represents the total number of times two vertices belong to the same community summed across all networks in the sequence. In short, the ``co-community graph'' records the relationships between community status across realizations. To assess vertex community stability in the graph sequence, we set a threshold $T$ where $t = \{1, \cdots, n\}$. As $t$ increments, we remove edge $(i, j)$ from $G_F$ if $t > w_{ij}$, so $G_F$ decomposes from a giant component and isolates emerge. Whenever a vertex becomes an isolate, we assign $t$ as a ``stability score'' attribute to this vertex. Intuitively, the larger the stability score, the more stable the vertex. When each vertex has a stability score, we can identify the most stable vertices in each community and use these for coloring communities (\autoref{sec:design-community-coloring}).

Our approach is designed specifically for sampled network realizations from a network distribution with a fixed vertex set. The Bernoulli sampling scheme causes network configurations to follow an approximately normal distribution by the central limit theorem, such that sampled realizations tend to have consistent structures and hence community memberships. However, this approach can be generalizable to other similar network sequences that follow a probability distribution.

\subsection{Visualizing the Network}
\subsubsection{Layout Engineering}
When arranging layouts for a graph sequence, too much emphasis on node stability across realizations diminishes readability, while optimizing individual layouts compromises analysts' ability to relate entities across states. NetHOPs address the trade-off between layout stability and readability \cite{brandes1997bayesian} by drawing on an aggregation and anchoring technique developed by \cite{brandes2012visualization}. This dynamic layout algorithm is based on the stress minimization approach and leverages the offline drawing property that all realizations in a sequence are known in advance. 

Given a graph $G = (V, E)$ defined by a vertex set $V$ and an edge set $E$, stress minimization computes a layout $P$ for $G$ using \autoref{eq:stress-mini}. To determine the positions $p_i$ for $i \in V$, $\delta_{ij}$ measures the dissimilarity, or the shortest path, between $(i, j)$ where $i, j \in V$ and $\|\cdot\|$ denotes the Euclidean norm. The weight matrix $W = \omega_{ij}$ determines the contribution of each $(i, j)$ to the layout arrangement.

\vspace*{-1mm}
\begin{equation}     
\label{eq:stress-mini}
    \text{stress}(P) =  \sum_{i < j} \omega_{ij} (\delta_{ij} - \| p_{i} - p_{j} \| )^2 
\end{equation}
\vspace*{-1mm}

Given a graph sequence $G^N$, we can compute a single reference layout $P$ through aggregation by \autoref{eq:aggregation}. The reference layout has the maximum stability as it places each vertex to a fixed position for each realization in $G^N$ (also called a ``flip-book'' approach~\cite{moody2005dynamic}). In \autoref{eq:aggregation}, $\bar{\delta}_{ij}$ is the mean shortest path for $(i, j)$ where $i, j \in V$. The weights $\omega_{ij}$ consider the variance of $\bar{\delta}_{ij}$, which places more importance on dyads having more stable shortest path length.

\vspace*{-1mm}
\begin{equation} \label{eq:aggregation}
    \begin{split}
        \text{st}&\text{ress}(P) = \sum_{i < j} \omega_{ij} (\bar{\delta}_{ij} - \| p_{i} - p_{j} \| )^2 \\
        &\text{where} \; \bar{\delta}_{ij} = \frac{1}{N} \sum^{N}_{n = 1} \delta^{n}_{ij} 
        \;\;\; \text{and} \;\;\; 
        \omega_{ij} = \frac{1}{\bar{\delta}_{ij}^2} \cdot \frac{1}{1 + \text{Var}(\delta_{ij})}
    \end{split}
\end{equation}
\vspace*{-1mm}

The reference layout, $P = p_{i}$ for $i \in V$, can be used as a benchmark to trade-off between layout stability and readability by \autoref{eq:anchoring-algorithm}.

\vspace*{-1mm}
\begin{align} \label{eq:anchoring-algorithm}
    \begin{split} 
        \text{stress}_{\alpha}(P^{n}) = & \underbrace{ (1 - \alpha) \cdot \; \sum_{i<j} \omega^{n}_{ij} (\delta^{n}_{ij} - \| p^{n}_{i} - p^{n}_{j} \| )^2 }_\textit{(characterizes individual layout difference)} + \\
                                   & \underbrace{ \alpha \cdot \; \sum_{i} \| p^{n}_{i} - p_{i} \|^2 }_\textit{(reference stability)} 
    \end{split}
\end{align}
\vspace*{-1mm}

Therefore, the balance between stability and readability is parameterized by $\alpha$. When $\alpha = 1$, all network realizations are positioned using $P$ obtained by \autoref{eq:aggregation}. On the other hand, setting $\alpha = 0$ removes the stability control term, so $P^n$ is computed by \autoref{eq:stress-mini}.

\subsubsection{Community Coloring}\label{sec:design-community-coloring}
Node color is perhaps the most common way to display community structure in node-link diagrams. Thus, we use the stability score (see \autoref{sec:design-community-matching}) and community labels to assign colors to communities. We reiterate the thresholding process used to compute the stability scores, and whenever we see a new component emerge, we assign a unique color from the Tableau 10 palette to the vertex with the highest stability score in the component if it does not yet have a color. When all stable vertices have colors assigned, we match the same colors to other vertices belonging to the same community in each network realization. The result of our matching algorithm is presented in \autoref{fig:community-coloring} (B). Additional marks, such as convex hulls, can be added to reinforce boundaries.

\subsection{Application of NetHOPs} \label{sec:prototype}
We describe an end-to-end application of NetHOPs used to create an interactive web-based prototype visualization system for study.

\textbf{Probabilistic Graph Model} Krackhardt's CSS data \cite{krackhardt1987cognitive} is a three-dimensional data structure used to measure individual perceptions of social relations within a network \cite{krackhardt1987cognitive, brands2013cognitive}. Following the notation used in \cite{krackhardt1987cognitive}, if $\mathscr{R}$ denotes a collection of adjacency matrices measuring a relation of interest, CSS can be represented as $\mathscr{R}_{i,j,k}$, where $i$ is the sender of a relation, $j$ is the receiver, and $k$ is the perceiver. For example, if $\mathscr{R}$ measures friendship, $\mathscr{R}_{1,2,3} = 1$ indicates person 3 perceives person 1 as a friend of person 2. We build our NetHOPs prototypes using two CSS datasets, which measure the advice-seeking and friendship relations among 21 managers in a high-tech firm. Therefore, our data has a dimension of $R \times N \times N \times N$ where $R = 2$ and $N = 21$. 

Traditionally, perceptions 
$\mathscr{R}^{\prime}_{i,j} = f(\mathscr{R}_{i,j,k_1}, \cdots, \mathscr{R}_{i,j,k_n})$
are used to build networks called \textit{consensus structures} through a dimension-reduction technique by \autoref{eq:css-aggregation} 
because the ground truth relations can be predicted from a weighted average of individual perceptions \cite{romney1986culture, freeman1987words}. However, \autoref{eq:css-aggregation} is essentially the thresholding approach to create networks (see \autoref{sec:background-probabilistic-graph}) because it adds an edge $(i, j)$ to the network if a certain proportion of perceivers claim $\mathscr{R}^{\prime}_{i,j} = 1$. 

\begin{equation}\label{eq:css-aggregation}
\mathscr{R}^{\prime}_{i,j} = 
    \begin{cases}
    1 & \text{if} \; \dfrac{1}{N} \sum_{k}\mathscr{R}^{\prime}_{i,j,k} \geq \; \text{Threshold} \\
    0 & \text{Otherwise}
    \end{cases}
\end{equation}

To avoid the thresholding approach, we create a weighted full graph $G = (V, E)$. A weight matrix $W = w_{i,j} \; \text{where} \; i \neq j \; \text{and} \;i, j \in V$ holds the aggregated perception of $(i, j)$ given by $k$ perceivers computed by $w_{i,j} = \frac{1}{N} \sum_{k}\mathscr{R}^{\prime}_{i,j,k}$. 
In reality, individuals can have varying degrees of perception accuracy depending on their experiences, observations, or acquaintances with each other. For demonstration purposes, we treat the quality of perceptions homogeneously, thus placing an equal weight when computing edge probabilities. 

Since the CSS data is a collection of perceptions provided independently by all individuals in the network, we treat each edge as an independent Bernoulli random variable. The edge variable $X_{i,j}$ can be directly modeled by its dyadic covariate $w_{i,j}$, and $Pr(X_{i,j} = 1) = E(X_{i,j}) = w_{i,j}$, which makes network sampling possible.
Each sampled realization in the sequence is a possible ``version of reality'' given the real-world social relations among the 21 managers. We repeatedly sample 150 network realizations for both the advice-seeking and friendship relations, and use these to create two sequences of 150 hypothetical outcomes, where the order in the sequence is not meaningful.

\textbf{Community Membership} For each sequence, we apply our community matching algorithm and recorded community membership.

\textbf{NetHOPs Rendering} We then compute a total of 11 sets of layouts using \autoref{eq:anchoring-algorithm} by setting the $\alpha$ parameter from zero to one with an increment of 0.1. Because of the offline drawing approach, users of the prototype can adjust layout stability and readability in the interface by tuning the amount of anchoring through a slider and receive immediate feedback. We then create another slider that enables users to adjust the animation speed in the unit of second per network, which ranges from 0.1 to 2. The interface also has four switches that allow users to control graph-specific visual elements, such as adjusting edge opacity between 0.2 and 1, or turn on or off visual aids like convex hulls, node color, and node labels designed specifically for community detection tasks.

\section{An Exploratory Study on NetHOPs}
We sought to investigate how well social network experts could use NetHOPs to conduct a broad range of common network analysis tasks, including density, path lengths, community detection, edge occurrence, and node attributes. We aimed to understand how well participants could process the uncertainty inherent in probabilistic graphs when estimating network properties or statistics. We used Krackhardt's CSS data (described in \autoref{sec:prototype}), which are relatively easy to describe for a study and include two graphs of varying connectivities. All experimental materials are included in the supplemental material.

\begin{table*}
\centering
\caption{NetHOPs user study tasks listed by task taxonomy with default visualization parameter value and graphical elements add-ons.}
\label{tab:nethops-tasks}
\vspace{-1mm}
\resizebox{\linewidth}{!}{
\begin{tabular}{clccccccc} 
\toprule
\multirow{2}{*}{\textbf{Taxonomy}} & \multicolumn{1}{c}{\multirow{2}{*}{\textbf{Task}}} & \multicolumn{1}{l}{\multirow{2}{*}{\textbf{Response Type}}} & \multicolumn{2}{c}{\textbf{Visualization Parameter}} & \multicolumn{4}{c}{\textbf{Graphical Elements}} \\ 
\cline{4-9}
 & \multicolumn{1}{c}{} & \multicolumn{1}{l}{} & \multicolumn{1}{l}{\textbf{Anchoring}} & \multicolumn{1}{l}{\textbf{Animation Speed}} & \multicolumn{1}{l}{\textbf{Dark Edges}} & \multicolumn{1}{l}{\textbf{Convex Hull}} & \multicolumn{1}{l}{\textbf{Node Color}} & \multicolumn{1}{l}{\textbf{Node Label}} \\ 
\hline
Topology & Estimate the number of distinct communities & Distribution & 0.8 & 1 & \cmark & \cmark & \cmark & \cmark \\
\multirow{2}{*}{Overview} & Estimate the number of isolates & Distribution & 0.8 & 1 & \cmark & \xmark & \xmark & \cmark \\
 & Estimate the graph density & Distribution & 0 & 1 & \cmark & \xmark & \xmark & \xmark \\
Browsing & Estimate the highlighted shortest path lengths & Distribution & 0.8 & 1 & \cmark & \xmark & \xmark & \cmark \\
Vertex-attribute & Estimate the vertex community stability & Probability & 1 & 0.5 & \xmark & \xmark & \cmark & \cmark \\
Edge-attribute & Estimate the occurrence of highlighted edges & Probability & 0.8 & 1 & \xmark & \xmark & \xmark & \cmark \\
\bottomrule
\end{tabular}
}
\vspace{-5mm}
\end{table*}

\subsection{Tasks}\label{sec:method-task}
Our tasks were inspired by the task taxonomies from Lee et al. \cite{lee2006task} and Ahn et al. \cite{ahn2013task}. We extended these tasks to focus on participants' ability to perceive the uncertainty of network statistics and properties. The available graph-specific objects in the CSS data imply tasks are centered around nodes, links, paths, and clusters (communities). \autoref{tab:nethops-tasks} presents the complete list of tasks we used, labeled according to \cite{lee2006task}. We omitted the isolate question on the advice-seeking CSS network because no isolates were detected after many sampling runs. 

Naturally, the extent to which we design custom parameterizations of the visualization to support specific tasks will affect performance. \autoref{tab:nethops-tasks} also lists the set of default visualization parameters and graphical elements we chose to provide for each task. We identified these parameter choices based on the target of the task. For graphical elements, this meant turning on visual features, like colors and convex hulls, helpful for community detection. Our goal was to provide minimal support so as to investigate how well analysts could do with a fairly stable visualization with extra visual features that would be easy to turn on or off (e.g., node or edge highlighting). For anchoring and animation speed, we used self-experimentation to choose parameter values that seemed most beneficial for the tasks. However, some of these choices can be arbitrary (e.g., node label, dark edges), so we designed a portion of our study to explore how analysts tuned the parameters themselves.

\subsection{Study Interface}
The study interface contained two columns, a visualization panel containing NetHOPs on the left, and a task panel showing questions on the right. The default view coupled each NetHOPs with an animation control panel, which allowed participants to pause, resume, forward play, or backward play realizations. When paused, participants could inspect NetHOPs' realizations one by one. A slider above the buttons indicated the progress of the animation, with a reset button next to the progress bar, which participants could click to reset to the first realization. The additional controls for NetHOPs' parameters mentioned in~\autoref{sec:prototype} were available for participants to use for the second part of the study, as we describe below.

\subsection{Response Elicitation}
Using a frequency-based distribution sketching tool called the distribution builder, our interface recorded probability estimates for node-attribute and edge-attribute tasks and elicited participants' uncertainty perception for topology, overview, and browsing tasks~\cite{sharpe2000distribution}. 

Distribution builders allowed participants to place a set number of balls (e.g., 20, 50, 100) in bins to express their beliefs about a parameter distribution. Prior work suggests this method leads to less noisy elicited beliefs than common approaches (e.g., asking for fractiles~\cite{goldstein2014lay} or sketching continuous density functions~\cite{hullman2017imagining}). 

For our distribution elicitation tasks, we requested participants first make deterministic estimates of the upper and lower bound of a network statistic. The interface then used the range produced by these estimates as the $x$-axis scale of the distribution builder. Participants could then drag different yellow markers above all possible discrete values that fall within the range, and allocate a total of 20 balls to each discrete ``bucket'' to approximate their perception of a distribution.  

\begin{figure}[tb]
    \centering
    \includegraphics[width=\columnwidth]{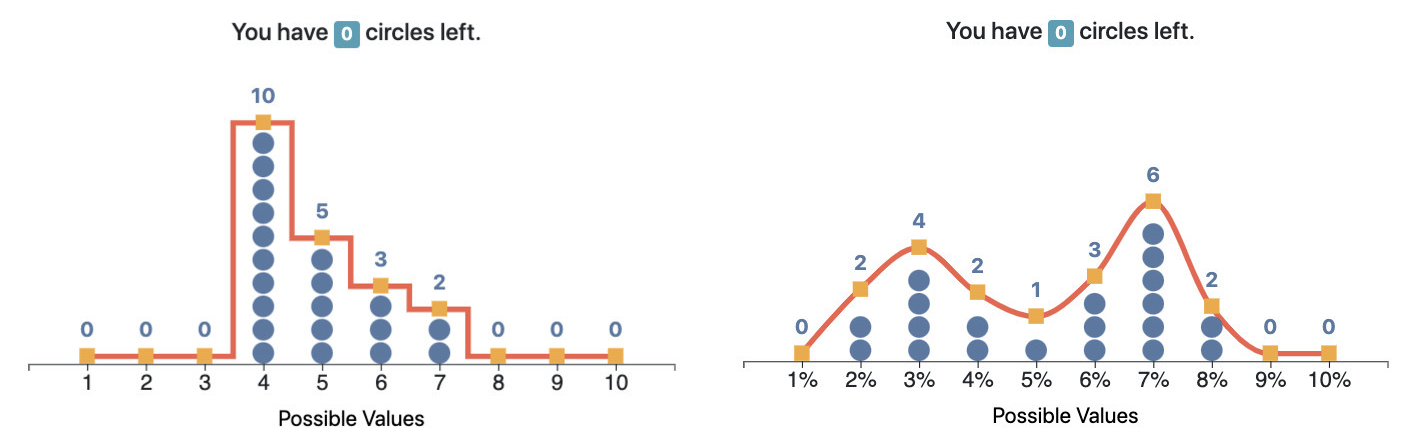}
    \setlength{\abovecaptionskip}{-3mm}
    \setlength{\belowcaptionskip}{-5mm}
    \caption{Modified distribution builder adapted from \cite{ahn2013task, hullman2017imagining} with kernel density curve used to elicit uncertainty perception.}
    \label{fig:distBuilder}
\end{figure}

Our study tasks required eliciting both discrete (e.g., path length) and continuous (i.e., network density) distributions. To help participants distinguish the difference, we created a modified distribution builder adapted from \cite{hullman2017imagining} by adding a kernel density curve as shown in \autoref{fig:distBuilder}. This instant feedback on the overall shape is important for continuous distributions because there are different ways a participant might try to use the interface. For example, if a participant wants to sketch a long-tailed distribution, she might forget to place balls in every other bin near the tail, which, unintentionally, results in a multi-modal distribution.

\subsection{Study Procedure}
Participants were directed to our web interface and instructed to complete the study in one session using Google Chrome. Participants were first required to pass a qualification test consisting of seven questions designed to check that they understood basic network statistics. The test questions were based on a simple static network diagram, and participants had to count or compute vertices, edges, isolates, communities, path length, and density, resembling the actual task questions shown in \autoref{tab:nethops-tasks}. To ensure participants were familiar with the use of distribution builders before tasks, we provided an exercise at the end of the qualification test. A bell-shaped normal distribution was given, and participants were asked to drag the yellow markers to mimic the distribution as much as possible. Participants passed the assessment if they made one or fewer errors, excluding the distribution builder exercise. If an error was detected, they were directed to a test results page, in which the system would show which question was wrong with detailed explanations of why and how to get the correct answer. 

Participants who passed the qualification test reviewed some background information, which described the CSS data structures, illustrated the NetHOP's data generating process, and reviewed community detection. At the end of the background section, participants were given detailed instructions on how to use the NetHOP's application interface and watched a short video visually demonstrating functionalities  with transcribed text shown on the same page. 

Participants began the tasks ordered by taxonomy shown in \autoref{tab:nethops-tasks}. They first completed the task on the advice-seeking NetHOPs and then proceeded to complete the same task using the friendship NetHOPs. Participants completed all tasks two times. In the first iteration, they completed the tasks using the default visualization parameters and graphical elements (\autoref{tab:nethops-tasks}). In the second iteration, they completed the same set of tasks but were given the freedom to tune the visualization parameters and control graphical elements. 

To ensure our participants fully understood how to use the visualization control panels to tune NetHOPs' displays, we created an instructional page describing the features of visualization parameters and graphical visual elements before proceeding to the second iteration. Another short video was created to provide a better visual aid explaining this newly added feature panel. In the second iteration, we reminded participants of each response they had previously provided to make it easier for them to tell if they thought the tuning could be improved. We instructed them to change their responses only if they were reasonably confident that tuning the visualization parameters and changing the graph elements could help them answer the task questions.

\subsection{Participants} 
We recruited participants familiar with SNA and graph theory by sending a recruitment email to a large SNA listserv and encouraged recipients to forward the study to relevant personnel. We also recruited participants from a private institution in the midwest, mainly from departments related to perception and networks (e.g., psychology, computer science, industrial engineering, and sociology). Participants who successfully completed the study received a \$15 dollar Amazon Gift Card or an equivalent for their time and effort completing the study. 

\subsection{Analysis Approach}
We computed the ground truth probabilities and distributions for all tasks using the 150 network realizations that comprised NetHOPs.

For tasks asking for deterministic probability estimates (node-attribute and edge-attribute tasks), we computed the differences between participants' guesses and the ground truth probability summarized from all realizations to assess how much they were off and in what direction.
For distribution elicitation tasks (overview, browsing, topology tasks), we discretized the ground truth distributions using the quantile dot plot (QDP) algorithm with pre-specified and constant bin width \cite{kay2016ish}. QDP is non-parametric and visualizes the predictive quantiles in Wilkinsonian dotplot \cite{wilkinson1999dot}, which allows us to match the ground truth distribution with the elicited distribution from the participant. 

To assess the differences between participants' elicited distributions and the discretized ground truth distributions, we used Earth Mover's Distance (EMD), which computes the minimum amount of work needed to shift the distribution of interest to match the target distribution of truth\cite{rubner1997earth,rubner1998metric}. QDP enables us to one-to-one point match the elicited distribution with the ground truth distribution formed by 20 balls; hence, the input distributions have the integer solution property, and each ball has the same weights of one \cite{hillier1995introduction}. 

Given a user input distribution $U = \{(u_1, w_1), \cdots, (u_i, w_i)\}$ and the ground truth distribution $T = \{(t_1, w_1), \cdots, (t_j, w_j)\}$, $|U| = |T| = N$ and $w_i = w_j = 1$ $\forall i,j \in N$. Let $F(U, T)$ be a set of all feasible flow to match balls in $U$ to $T$, and so the distance or cost to perform such movements is the Manhattan distances, $L_M(u_i, t_j)$, summed across all pair of balls, which can be expressed by $Cost(F, u, t) = \sum_i \sum_j f_{ij}L_{M}(u_i, t_j)$. Therefore, the EMD score can be computed by \autoref{eq:emd}, which is an optimization aiming to minimize the cost of flows. EMD score can be interpreted as the average distance balls in the user distribution must move to match those of the target distribution.

\begin{equation} \label{eq:emd}
    EMD(U, T) = \frac{min_{F = (f_{ij}) \in F(u, t)} Cost(F, U, T)}{N}
\end{equation}

Below, we report mean estimates with bootstrapped 95\% confidence intervals (CIs) of participants' responses and EMDs for each task.

\section{Results}
\begin{figure*}[ht]
 \centering 
 \includegraphics[width=\textwidth, keepaspectratio]{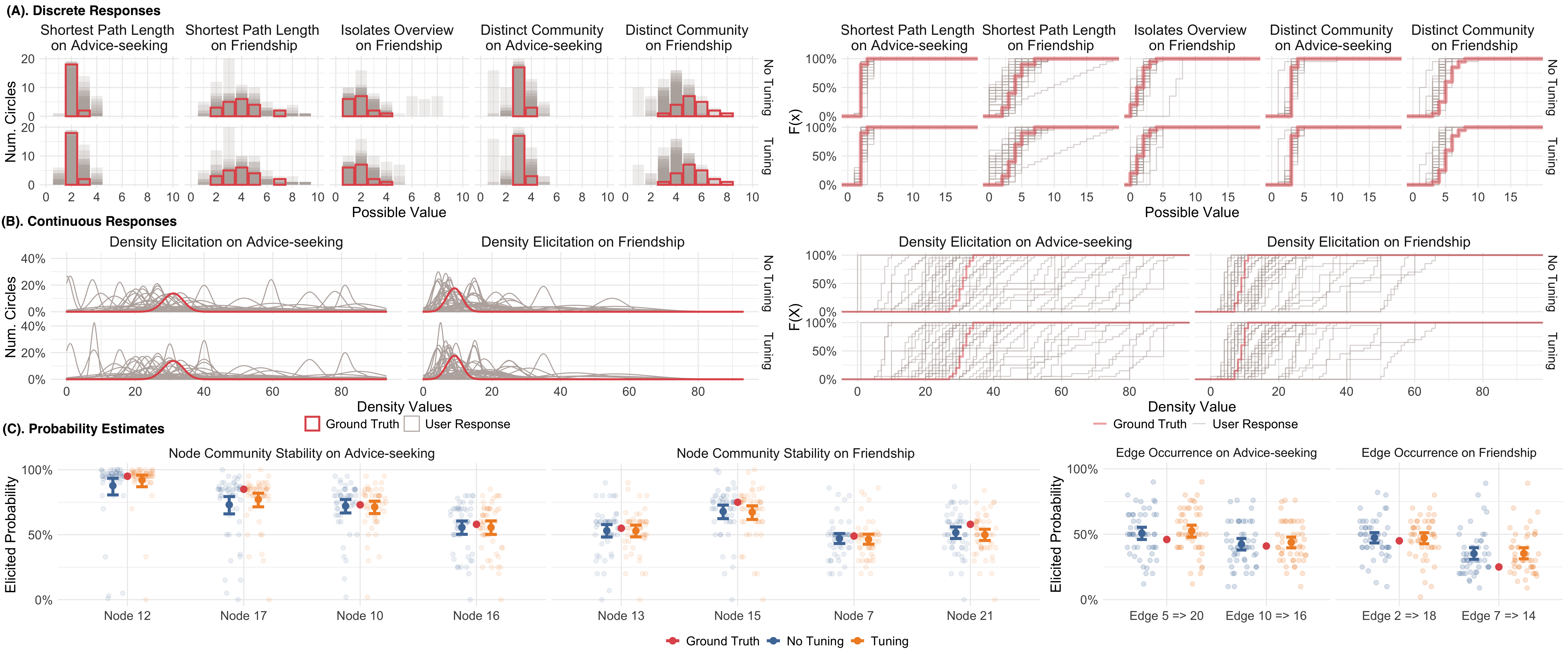}
 \setlength{\abovecaptionskip}{-3mm}
 \setlength{\belowcaptionskip}{-3mm}
 \caption{
 (A). Left: Barplots display participants' discrete responses for browsing, topology, and overview (isolate) tasks stacked on top of each other against the ground truth distribution in red. Right: Empirical cumulative distribution functions based on participants' elicited discrete responses shown in the barplots on the left.
 (B). Left: the density plot shows participants' continuous responses for the network density task stacked on top of each other against the ground truth distribution in red. Right: Empirical cumulative distribution functions based on participants' elicited network density estimations are shown in the density plot on the left. 
 (C). Strip plots present participants' deterministic probability estimates with 95\% bootstrapped CIs for each attribute-based task from the study. The ground truth probabilities for each task are shown as red points between the CIs. We elicited four probability estimates for node-attribute tasks and two for edge-attribute tasks from both advice-seeking and friendship networks.
 }
 \label{fig:user-response}
\end{figure*}

\subsection{Preliminaries}
Of 173 participants recruited to our study website, 32\% (56) passed the qualification test. The high exclusion rate indicates our stringent requirement on the expertise of the participants. We removed five entries due to incomplete data, leaving 51 responses for analysis. The majority of our participants are graduate-level network researchers (98\%) who have taken at least one course in SNA or graph theory (90\%) and use network analysis in their daily work (86\%). 

We provide an overview of performance in \autoref{fig:user-response}, with responses superimposed against the ground truth distribution (rows A and B) or probability (row C). Empirical cumulative distributions functions (A and B right) are plotted for all distribution elicitation tasks. Barplots and density curves differentiate the discrete (A left) and continuous (B left) responses. We present results from the first block of trials where visualization parameters were given separately from those for the second block where tuning was allowed.

In reviewing overall performance, we discovered a data anomaly for the browsing task of the friendship network in \autoref{fig:user-response}. About half of participants (22) incorrectly assumed some realizations had a shortest path length of zero (Figure~\ref{fig:user-performance} A, left, second column) when theoretically it should be infinity and therefore not counted when sketching the distributions. This conceptual misunderstanding led to more divergence (i.e., higher EMD scores) in response quality for this task.

\subsection{Performance without Tuning}
\begin{figure}[tb]
 \centering 
 \includegraphics[width=\columnwidth, keepaspectratio]{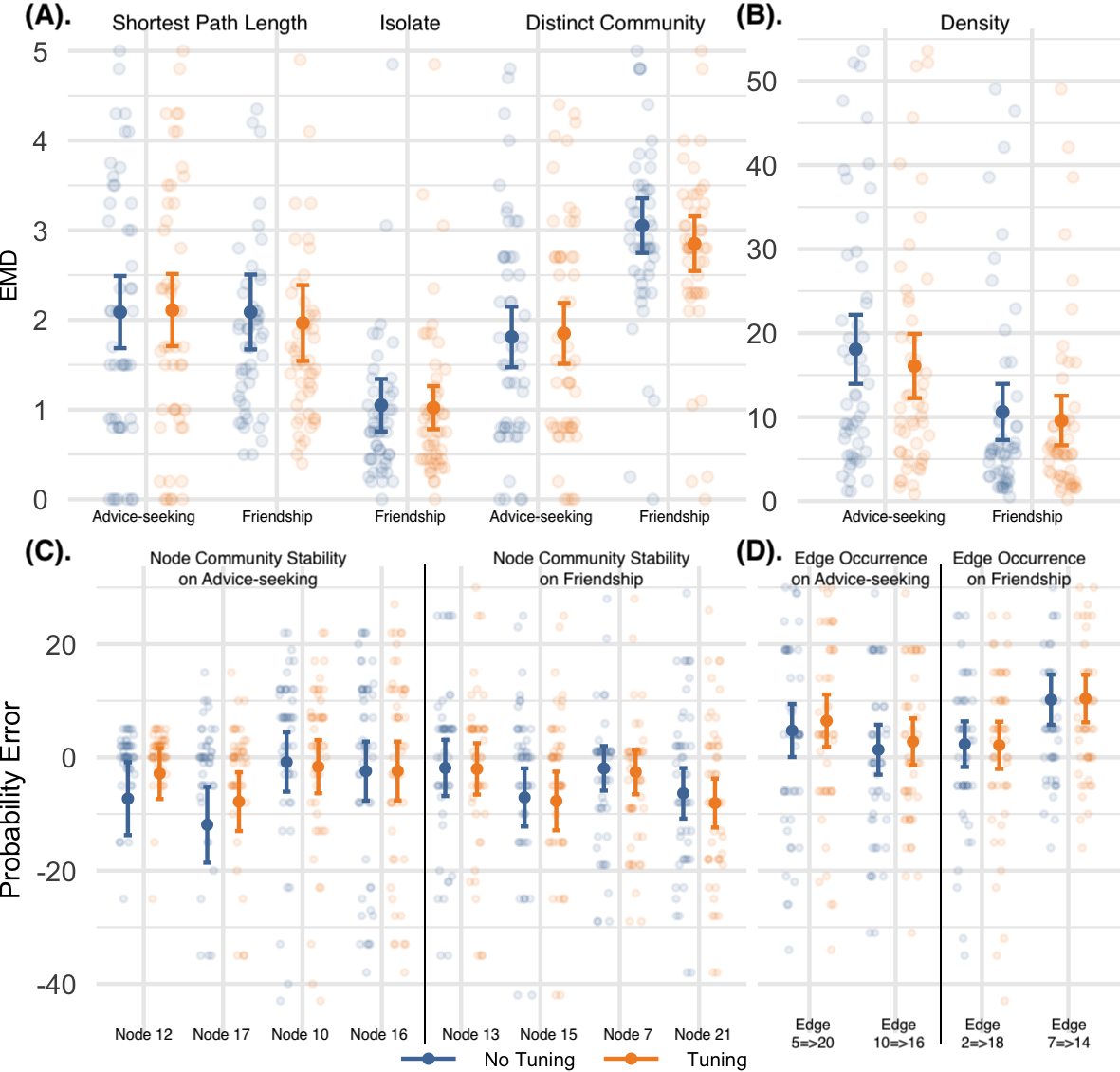}
 \setlength{\abovecaptionskip}{-3mm}
 \setlength{\belowcaptionskip}{-6mm}
 \caption{ 
 (A) and (B) display the bootstrapped mean EMD scores with 95\% CI for browsing (shortest path length), overview (isolate and network density), and topology (distinct community) tasks. 
    (C) and (D) display the bootstrapped mean of probability estimation errors for each attribute-based task on node community stability and edge occurrence.
 }
 \label{fig:user-performance}
\end{figure}

We evaluated individual task performance by computing the bootstrapped mean EMD scores for distribution elicitation tasks and the bootstrapped mean estimation error for the probability estimation tasks. The blue and orange 95\% CIs in \autoref{fig:user-performance} indicate tasks completed with or without tuning correspondingly.

\textbf{Browsing Task}
Participants performed consistently well when tracking the shortest path lengths between selected vertices on both the advice-seeking and the friendship networks, despite the smaller variance for the ground truth distribution in the advice-seeking network than that of the friendship network shown in \autoref{fig:user-response} (A). Participants were more likely to overestimate the shortest path length for the advice-seeking network, and, as described above, underestimate for friendship.  

Despite these small differences in apparent bias in the results, the two blue CIs in \autoref{fig:user-performance} (A) for the two browsing tasks almost completely overlap. The mean EMD for advice-seeking has a CI $[1.69, 2.49]$ compared to a CI $[1.67, 2.5]$ for the mean EMD of friendship.

\textbf{Overview Tasks}: 
Most participants excelled at the isolate identification task, with the response distributions closely resembling the ground truth (\autoref{fig:user-response} A, middle columns). The blue CIs for the isolate task in the middle of \autoref{fig:user-performance} (A), top, come closest to zero EMD. Network density estimation, on the other hand, was the most difficult task for participants. Density plots and CDFs in \autoref{fig:user-response} (B) show lots of dispersion in elicited distributions, with some centered roughly 40\% points away from the ground truth distribution's location. The mean EMD scores for the density estimation tasks are much higher than the rest of the distribution-elicitation tasks in \autoref{fig:user-performance} (B). Participants, on average, perceived the density distribution more accurately on the sparse friendship network (CI$[7.3, 13.9]$) than that of the advice-seeking (CI$[13.9, 22.2]$). The difficulty of density estimation was also reflected in comments that participants left at the end of the study (e.g., ``Density seems near impossible to intuitively estimate'' and ``I was able to count for the first few questions but that is clearly not practical for the density ones.'').

\textbf{Topology Task}
Results for the topology tasks counting distinct communities suggest participants perceived the distributions more accurately in the advice-seeking network. The elicited distributions are clearly biased toward underestimation for the sparse friendship network (\autoref{fig:user-response} A, last column) but less clearly biased and closer to the ground truth for advice-seeking. The blue CIs on EMD in \autoref{fig:user-performance} (row A) do not overlap for these tasks (advice-seeking: $[1.47, 2.14]$, friendship: CI$[2.75, 3.36]$). This outcome is unsurprising because the ground truth distribution for the advice-seeking networks shown in \autoref{fig:user-response} (A) has a smaller variance compared with that of the friendship network. 

\textbf{Attribute-based Tasks} 
\autoref{fig:user-performance} (C) and (D) display how much participants' probability estimates were off from the ground truth for the attribute-based tasks. Participants tended to underestimate probabilities for node-attribute tasks and overestimate probabilities for edge-attribute tasks. However, on average, participants' probability estimates were not too far from the ground truth. As suggested by the blue CIs in \autoref{fig:user-performance} (C) and (D), all participants' deterministic probability guesses were off within 20 percentage points of the ground truth probability.

\subsection{Performance with Tuning}
Several participants commented that the ability to control NetHOP rendering helped them complete the tasks. We summed each participants' EMD scores for all tasks completed with and without tuning and computed the average amount of improvement. We found that the ability to control NetHOPs rendering could improve participants' distribution elicitation by 4\% on average, though this was not a reliable difference given our sample size (CI $[-4.2\%, 12\%]$). We computed a similar statistic by aggregating the total absolute error for probability estimation tasks, and found tuning visualization parameters did not improve probability estimation (CI$[-14.2\%, 13.3\%]$).

Recall that we instructed participants to update their answers only if they were confident that tuning could help them achieve better results. Participants performed reasonably well with the default parameters, which makes this result not terribly surprising. Many may not have felt they could improve the prior parameters, or tweaked them only by a small amount. This is reflected in that 22\% (11) of our participants provided the same answers in the second iteration of tasks. Nonetheless, we provide an exploratory analysis of tuning strategies, since how participants changed parameters can shed light on how visualization parameters may support different tasks in slightly different ways. 

\subsubsection{Tuning Strategies}
\begin{figure}[tb]
 \centering 
 \includegraphics[width=\columnwidth, keepaspectratio]{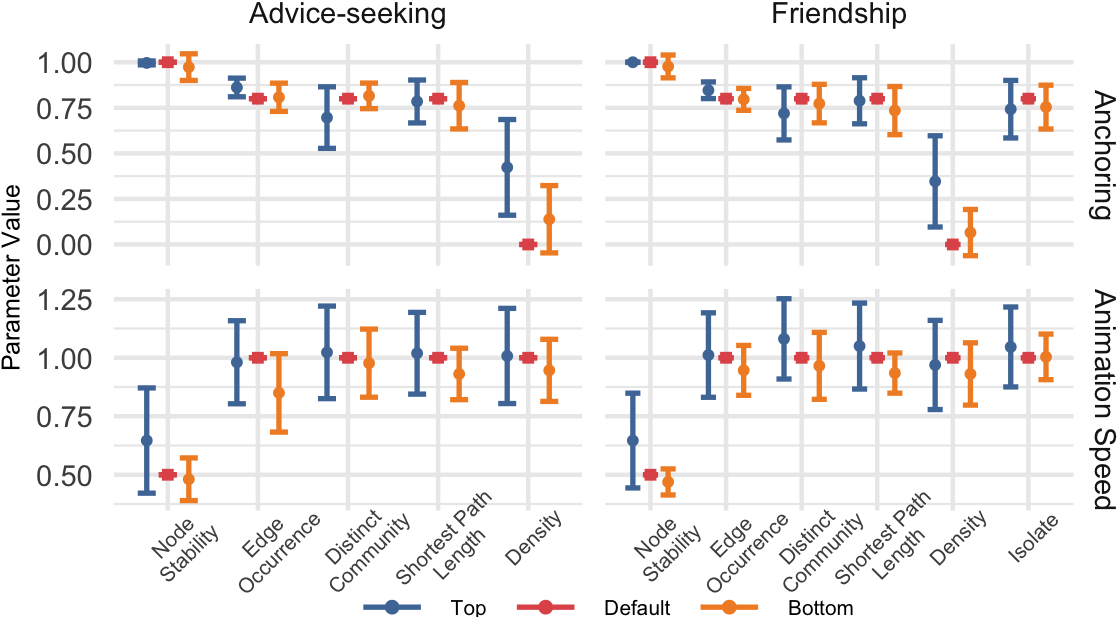}
 \setlength{\abovecaptionskip}{-3mm}
 \setlength{\belowcaptionskip}{-5mm}
 \caption{ 
 Bootstrapped means with 95\% CIs for anchoring (A) and frame rate (B) parameters used by participants.
 }
 \label{fig:slider-CI}
\end{figure}

The distributions of the two performance metrics indicate that tuning did help some participants. To investigate the dynamics between visualization parameters and accuracy, we ranked all participants by their performance with tuning and grouped top-performers from the first quartile and bottom-performers from the fourth quartile, with each group consisting of 13 participants. Top-performers had a pooled mean EMD score of 3.1 (Min.: 1.7, Median: 2.8, SD: 1.2, Max.: 5.3), and their probability estimates were, on average, 1.9\% off from the truth (Min.: 0.1, Median: 1.3, SD: 1.6, Max.: 5). On the other hand, bottom-performers had a pooled mean EMD score of 7.8 (Min.: 3.1, Median: 6.3, SD: 4.7, Max.: 15.4), and their probability estimates were, on average, 14.3\% off from the truth (Min.: 1.3, Median: 12.6, SD: 12.1, Max.: 44.9). We focused on how anchoring and animation speed were used by each group, and present the bootstrapped mean with 95\% CIs in \autoref{fig:slider-CI} to assess how the parameters related to response quality.

\textbf{Tuning on Layout Stability}
\autoref{fig:slider-CI} row (A) shows participants' anchoring tuning did not deviate too much from the default value we provided for most tasks, except for density estimation. For this overview task, participants preferred layout stability over the general readability the layout algorithm optimizes for (e.g., by reducing overlapping edges). This goes against our expectation that the need to estimate density would trump the need to keep nodes in consistent positions, which had led us to set a default anchoring value of zero. 

Layout stability appears more important for the two attribute-based tasks. Participants preferred higher average anchoring for node-attribute tasks regardless of performance, and nearly all top-performers chose the most stable layout for node-attribute tasks.

The topology tasks (detecting distinct communities) and the isolate overview task were the only two tasks where top-performers on average preferred lower anchoring values than the bottom-performers. Anchoring preferences were similar between networks for the browsing tasks. Both groups lowered the anchoring slightly from the default of 0.8, but the bottom-performers, on average, lowered anchoring a bit more.

We pooled all anchoring parameters used by top-performers and bottom-performers for each task and found that top-performers preferred layout stability more than bottom-performers. Top-performers tended to set a slightly higher average anchoring of 0.81 (CI $[0.73, 0.89]$) compared with the average 0.71 (CI $[0.63, 0.79]$) set by the bottom-performers. Between CSS datasets, participants preferred more layout stability when working on the denser advice-seeking network. Top-performers used an anchoring of 0.83 (CI $[0.75, 0.90]$) on average for the advice-seeking and 0.78 (CI $[0.70, 0.86]$) for the friendship. This penchant is more noticeable from bottom-performers as the averaged anchoring used on the advice-seeking network is 0.72 (CI$[0.63, 0.81]$), and the friendship network is 0.67 (CI$[0.59, 0.75]$).

\textbf{Tuning on Animation Speed}
One consistent pattern observable from \autoref{fig:user-performance} (B) is that top-performers generally chose to set the animation speed slower than the bottom-performers. Top-performers, on average, preferred to render each network for 1 second (CI $[0.9, 1.13]$), compared to the average 0.82 (CI $[0.75, 0.9]$) of bottom-performers.

When viewing the denser advice-seeking networks, all participants from both groups preferred faster animation speed on average, displaying each network realization on the screen for 0.77 seconds on average (CI $[0.66, 0.88]$), compared to 0.84 seconds (CI $[0.75, 0.92]$) for the friendship network. For top-performers, the averaged animation speed when completing tasks on the advice-seeking network was 0.97 seconds per network (CI$[0.85, 1.1]$), and 1.02 seconds per network (CI$[0.91, 1.13]$) for the friendship network. Similarly, bottom-performers chose to display each network realization on the screen for 0.77 seconds (CI$[0.68, 0.86]$) for the advice-seeking network and 0.83 (CI$[0.76, 0.90]$) seconds for the friendship network. 

\textbf{Graphical Elements}
Edge opacity, node color, convex hulls, and node labels are network-related graphical attributes or elements that can be highly task-specific. In analysis, we found it hard to generalize much from looking at each individually, but analyzing combinations of visual parameters was more meaningful.

\begin{figure}[tb]
 \centering 
 \includegraphics[width=\columnwidth, keepaspectratio]{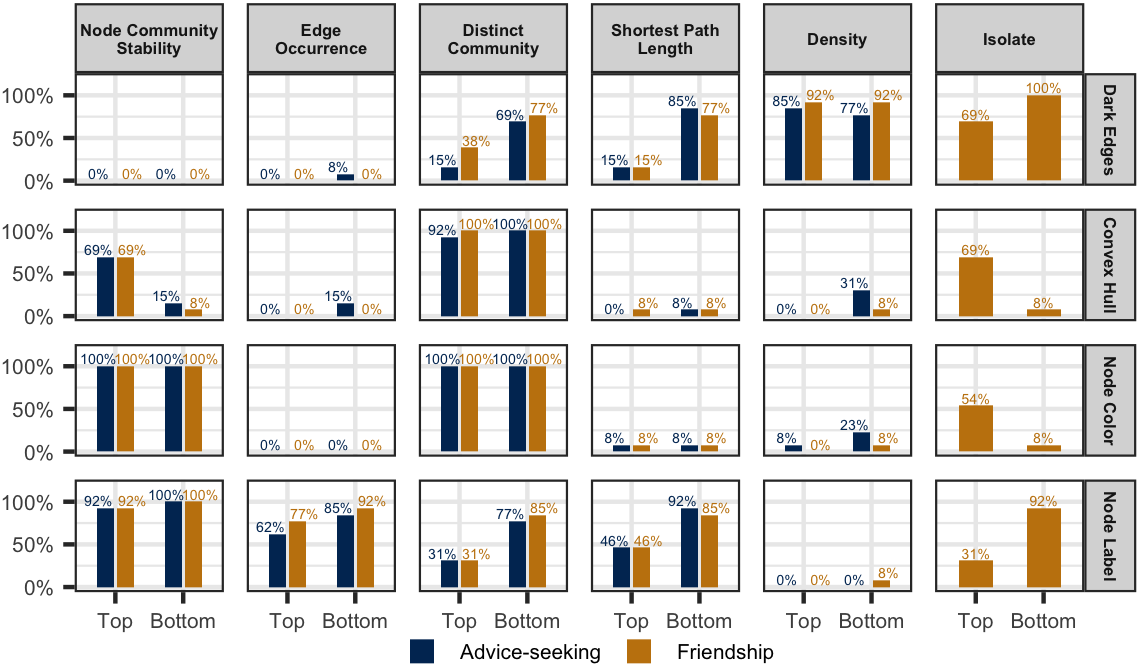}
 \setlength{\abovecaptionskip}{-3mm}
 \setlength{\belowcaptionskip}{-5mm}
 \caption{ 
 Participants' use of graphical visual aids.
 }
 \label{fig:switcher-param}
\end{figure}

We computed the percentage of participants who used each graphical element by task in \autoref{fig:switcher-param} and found some combinations were very different from the default visualization that we thought would work best on a given task. The majority of top-performers chose to make targeted or relevant network objects and elements more salient by deactivating irrelevant visual elements. For example, when detecting the number of communities, they chose to make edges less salient and turn off node labels to better emphasize node color and convex hulls. When browsing shortest paths, 75\% of top-performers preferred less salient edges to further emphasize the red highlighted edges connecting the nodes. To estimate network densities, top-performers preferred more salient edges while minimizing other visual elements by turning off convex hulls, node color, and node labels.

We also observed some creative use of visual features. For example, when identifying the number of isolates, roughly 70\% of top-performers turned on convex hulls, 50\% removed node color, and 30\% preferred less salient edges, even though convex hulls and node colors were seemingly irrelevant visual elements to the task. We note that convex hulls do not include isolated nodes, and our coloring algorithm would consistently assign a distinct color to isolated nodes. With these two parameters, edge salience becomes less important and could be turned off to accentuate convex hulls and node color.

\subsection{Precision of Inference \& Time-Accuracy Correlation }\label{sec:precision-correlation}
Two forms of error can impact the precision of inferences about network statistics made with NetHOPs: perceptual and cognitive errors related to how accurately analysts can estimate probabilities from the visualization (which applies to any visualizations of distributions), and approximation error introduced by sampling from the network model.

\begin{figure}[tb]
 \centering 
 \includegraphics[width=\columnwidth, keepaspectratio]{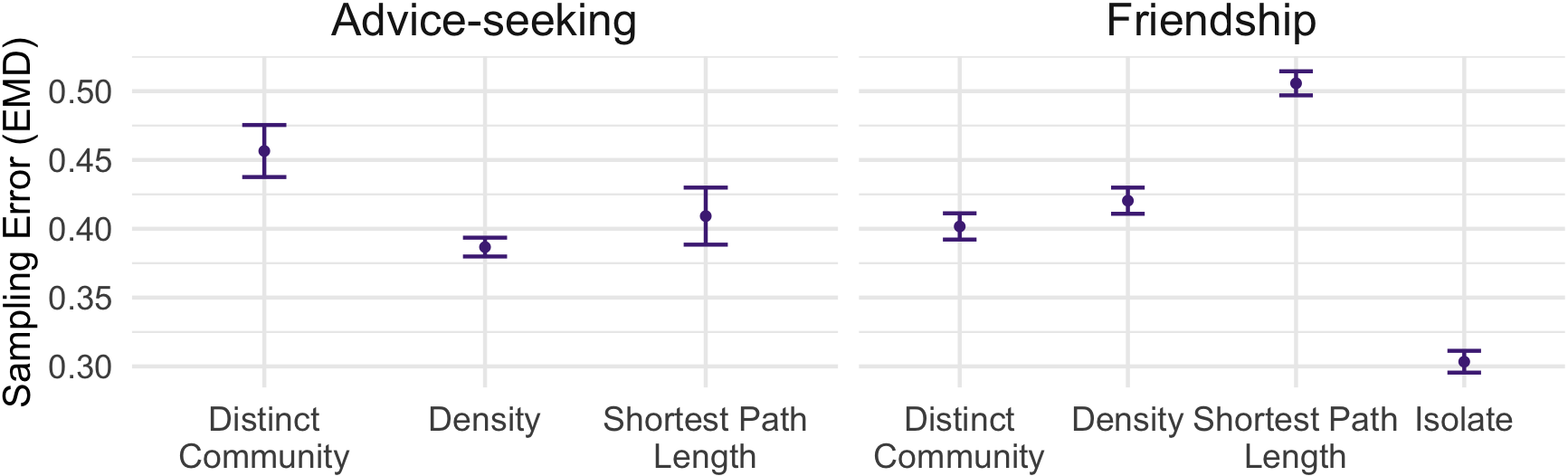}
 \setlength{\abovecaptionskip}{-2mm}
 \setlength{\belowcaptionskip}{-5mm}
 \caption{ 
 Sampling error for the distribution elicitation tasks in EMD based on 500 samples of 150 network realizations.
 }
 \label{fig:EMD-sampling-error}
\end{figure}

First, we quantify the sampling error introduced when taking a set of random draws from the graph model. While we cannot compute approximation error against the ground truth model, we can infer it by re-sampling $N$ sets of network realizations from the model and using the distributions of network statistics from each re-sampled set to compute EMD scores against those from the set of NetHOPs. We can then quantify the sampling error of the distribution for each network statistic in the unit of EMD by constructing a confidence interval, as shown in \autoref{fig:EMD-sampling-error}. Therefore, if a participant perfectly perceived and sketched a distribution and received an EMD score of zero using NetHOPs, her perception could be off by an $\mu_{EMD} \pm SE_{EMD}$ amount. This approach is generalizable and can be easily applied to compute the sampling error of any probability estimation task.

Naturally, participants may not view all NetHOPs' realizations. While we did not log exact realizations viewed, we can use time spent on tasks to infer approximate viewing. For our study, we found participants, on average, spent 55 minutes to complete all tasks (Min.: 9, Median: 49, SD: 32, Max.: 176), after removing one outlier of 549 minutes. This participant paused on one task page for approximately 8.5 hours, suggesting this participant might have left the browser open and walked away. Recall that the average of 55 minutes describes participants' time to complete a total of 70 questions spread across 22 task screens. For the distribution elicitation tasks, 75\% of participants completed the tasks within approximately five minutes, which includes the time spent on identifying the lower and upper bounds and sketching the distributions. For probability elicitation tasks, 75\% of our participants were able to complete all four sub-tasks for node stability within six minutes and two sub-tasks for edge occurrence within three minutes. These estimates suggest that many participants viewed all 150 frames, though we cannot be sure how they divided their attention between questions and watching the animation.

\begin{figure}[tb]
 \centering 
 \includegraphics[width=\columnwidth, keepaspectratio]{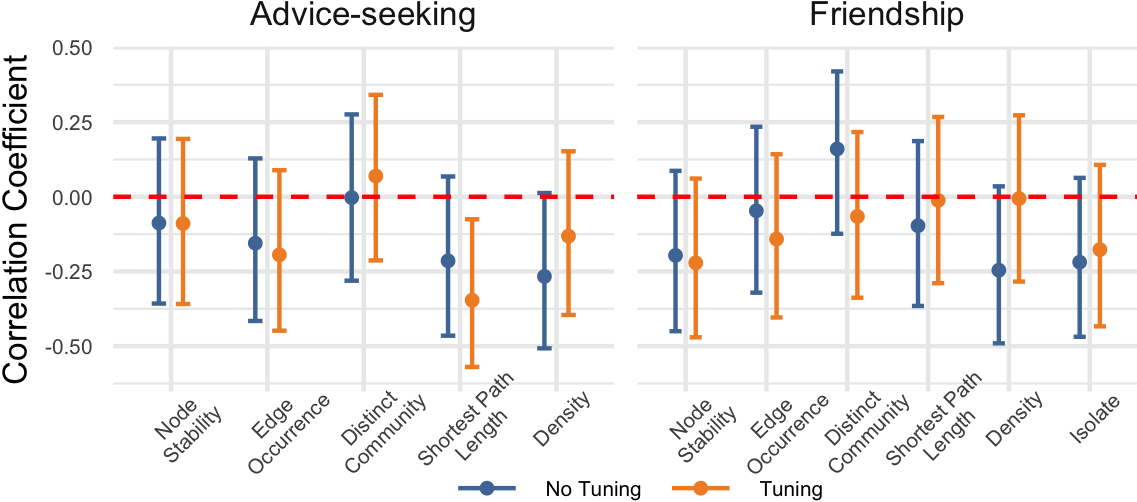}
 \setlength{\abovecaptionskip}{-2mm}
 \setlength{\belowcaptionskip}{-5mm}
 \caption{ 
 Pearson's Correlation with 95\% CIs between task completion time (mins.) and perception accuracy.
 }
 \label{fig:correlation-CI}
\end{figure}

To assess whether spending more time on a task correlated with performance, we conducted a correlation analysis between task completion time and response quality by computing Pearson's correlation coefficient for each task, then by using Fisher’s Z-Transformation~\cite{fisher1915frequency, fisher1921probable} to meet the normality assumption and compute CIs shown in \autoref{fig:correlation-CI}. We observed negative correlation coefficients in the range of $[-0.3, 0]$ for the majority of the tasks, implying that longer time spent on tasks can slightly reduce the amount of perception error. However, almost all CIs include zero, which makes this effect unreliable except for the task on the shortest path length of the advice-seeking network with tuning. It is worth mentioning that participants left comments such as, ``I had to think a lot about each question and pause to take breaks. There were a lot of questions,'' which may bias these correlations.

\section{Discussion}
Our user study demonstrates NetHOPs' potential to support exploratory analysis on probabilistic networks. We summarize our findings, provide design recommendations, and discuss future work. 

\textbf{Overall Performance}
When using NetHOPs, we discovered analysts were better at identifying isolates, tracking path lengths, and detecting attribute-based changes on both nodes and edges, while they struggled more with estimating properties like density and distinct communities when realizations had overlapping communities. We speculate the reason our participants excelled at identifying isolates and tracing paths is that people tend to be visually sensitive to a lack of continuity between nodes, perhaps because it can be recognized pre-attentively when layouts are stable. 

We found that participants' estimates of network properties fell on average within 11\% of the statistics computed on the realizations comprising NetHOPs. Naturally, a question arises of whether this amount of error, in combination with the sampling error discussed in \autoref{sec:precision-correlation}, is tolerable. We suspect that the error induced by NetHOPs is acceptable in many cases where visual analysis aims to help analysts to get a rough sense of many network properties to build intuition. We also note that whenever visualization is used for network exploration, a rough sense of the properties is more likely to be the goal, since an analyst might otherwise use more exact methods or modeling. However, future work might better contextualize whether the approximations allowed by NetHOPs seem sufficient to network analysis. 

\textbf{Layout Stability}
Our participants appeared to tune NetHOPs differently based on network density. Stable layouts seemed generally beneficial for all tasks, even for tasks like density estimation that do not require stability. A highly stable layout may prevent change-blindness and make it easier to detect important differences across realizations. Interestingly, our study participants did not prefer a completely stable layout except for attribute-based tasks on nodes. This suggests that estimating distribution from animated network realizations requires its own balance of stability versus within-realization optimization as analysts appear sensitive to both. 

We also found that analysts can tolerate more node movements for sparse networks. We speculate the reason behind this may be related to the overall shape of the networks and the amount of visual clutter in the visualizations caused by density. A sparse network is more visually trackable because of fewer graphical elements, so stability can be compromised in favor of readability. For denser networks, we think methods, such as \cite{nocaj2014untangling}, can be incorporated into the layout engineering when certain network properties (e.g., small-worldness) are known to make layout more comprehensible before aggregation and anchoring.

\textbf{Animation Speed}
We found top-performers tended to use slower animation speeds, which is unsurprising as a slower speed allows a more thorough review of each network realization. However, a faster animation speed (and a high degree of anchoring) appears to work well for node-attribute tasks where change detection is straightforward. The chosen animation speeds for the two NetHOPs in our user study are relatively slower compared to the chosen default speeds for simpler 2D HOPs in prior work~\cite{hullman2015hypothetical,kale2018hypothetical,kale2020visual}. This is because network realizations convey more information than simpler charts and require more cognitive load for analysts to process. Participants may have needed a longer time to register what relationships were present in a realization.

\textbf{Graphical Elements}
The salience of relevant network objects appears to be the most important driver of participants' choices. Therefore, graphical aids like edge opacity, node color, convex hulls, and node labels are useful in producing more accurate uncertainty depictions. Highlighting or emphasizing relevant network objects can also help analysts identify targeted graphical elements and capture changes with minimum effort. We think these add-ons can greatly reduce the risks of change-blindness, and suggest any practical implementation of NetHOPs would allow for such visual tuning. 

\textbf{Limitations and Future Work}
NetHOPs are subject to errors from (1) approximating a distribution with samples to create the animation, and (2) analysts not watching full realizations. As we described in \autoref{sec:precision-correlation}, errors can be quantified by comparing network statistic distributions across sets of realizations of a particular size. However, future work might aim to better understand how many network realizations analysts tend to watch before they feel confident estimating a particular statistic, providing more insight into how much time is required and for how much accuracy gain, in relation to static graphs. Our study also involved two relatively small networks. Beyond testing larger networks, researchers could explore the utility of NetHOPs for bipartite or multiplex networks, as well as network models with edge dependencies, which NetHOPs are well suited to address. 

\section{Conclusion}
NetHOPs aim to facilitate uncertainty communication for probabilistic networks. We summarized the design of NetHOPs and illustrated the computation of its controllable dynamic layout. We designed a community matching algorithm so our technique can support uncertainty detection of topology-based tasks such as clustering identification. The results of our user study suggested that the technique can support visual exploratory analysis, at least of small networks. Our results point to directions for future work around optimizing visualization parameters and better understanding perception of network properties via animation.

\acknowledgments{
This work was supported by NSF IIS-1907941, NSF IIS-1815760, and Microsoft.
}

\bibliographystyle{abbrv-doi}

\bibliography{references}
\end{document}